\documentclass[nofootinbib,12pt]{article}
\usepackage{graphicx}

\begin{document}

%\newcommand{\dlt}{\bigtriangleup}
%\newcommand{\beq}{\begin{equation}}
%\newcommand{\eeq}[1]{\label{#1} \end{equation}}
%\newcommand{\insertplot}[1]{\centerline{\psfig{figure={#1},width=15.0cm}}}
%\newcommand{\insertplotsshort}[1]{\centerline{\psfig{figure={#1},height=5.0cm}}}
%\newcommand{\insertplotll}[1]{\centerline{\psfig{figure={#1},height=11.0cm}}}
%\newcommand{\insertplotlll}[1]{\centerline{\psfig{figure={#1},height=18.0cm}}}

%\parskip=0.3cm

%\begin{titlepage}

\vskip 0.5cm \centerline{\bf Which Pomeron survives at LHC energies?} \vskip 0.3cm
%\centerline{V.~Bitev$^{\star}$, A.~Lengyel$^{\diamond}$ and Z.~Tarics$^{\diamond}$}
\centerline{A.~Lengyel and Z.~Tarics}

\vskip 0.1cm
%\centerline{  Institute of Electron Physics,}
%\centerline {Academy of Sciences of Ukraine, Uzhgorod, 88017, Ukraine}
%\centerline{alexander-lengyel@rambler.ru}\
%\centerline{iep@iep.uzhgorod.ua}
 \vskip 0.1cm

%\centerline{ $^\star$ \sl JINR, Dubna, 140980,Russia}
%\centerline{ $^\diamond$ \sl Institute of Electron Physics,}
%\centerline {Academy of Sciences of Ukraine, Uzhgorod, 88017, Ukraine}
%\centerline{ $^\diamond$ \sl IEP, Uzhgorod, 88017, Ukraine}
\centerline{Institute of Electron Physics, Uzhgorod, 88017, Ukraine}
\centerline{alexander-lengyel@rambler.ru}\
\centerline{tarics1@rambler.ru}
% \vskip 0.1cm
%\author{V.~Bytev\,$^c$}\email{bytev@thsun1.jinr.ru}
%\centerline\author{A.~Lengyel\,$^d$}\email{alexander-lengyel@rambler.ru}\\
%\centerline\author{Z.~Tarics\,$^d$}\email{iep@iep.uzhgorod.ua}
%\centerline{alexander-lengyel@rambler.ru}\
%\centerline{iep@iep.uzhgorod.ua}
 \vskip 1cm

%$^c$Bogolubov Laboratory of Theoretical Physics, Joint Institute for Nuclear Research, Dubna, %140980 Russia\\
 %$^d$Institute of Electron Physics, National Academy of Sciences of Ukraine, Uzhgorod, 88017 Ukraine}
\begin{abstract}
An eikonalized elastic proton-proton and proton-antiproton scattering amplitude $F(s,t)$, based on the suggestion of a finite sum of ladder diagrams, calculated from QCD and the number of $s-$channel gluon rungs and correspondingly the powers of logarithms in total cross section depends on available increasing energy. Explicit expressions for total cross section involving three and four rungs (four and five prongs) with $ln^3\left(s\right)$ and   $ln^4\left(s\right)$ as highest terms, respectively) are fitted to the all available proton-proton and proton-antiproton total cross section data. Predictions for $pp$ total cross section at LHC energy are given and compared with prediction of several Regge-models including possible hard Pomeron contribution.
\end{abstract}

%\pacs{11.80.Fv, 12.40Ss, 13.85Kf}

%\maketitle

%\vskip 0.1cm

%\vfill

%$\begin{array}{ll}
%^{\star}\mbox{{\it e-mail address:}} & \mbox{mario.deile@cern.ch}\\
%^{\diamond}\mbox{{\it e-mail address:}} &
%  \mbox{jenk@bitp.kiev.ua}

%$\end{array}

%\end{titlepage}

\section{Introduction}
The renewed interest in high-energy soft interactions is largely triggered  by the recent LHC measurements by the TOTEM \cite{7tev}
%,CMS \cite{CMS} and ATLAS \cite{ATLAS}
measurements of $pp$ scattering. The accessible energies
offer new possibilities to check the basic theoretical concepts, such as unitarity and the properties of the (multy) Pomeron exchanges.

In the present paper we develop further and calculate the effects coming from the inclusion of both a "soft" and "hard" Pomeron, the Pomeron
being a finite sum of reggeized gluon ladders, different from the "QCD Pomeron", which is an infinite sum of such ladders~\cite{ FKL,BL,L}, resulting in the so-called supercritical behavior of the total cross section. In that approach, the main contribution to the inelastic amplitude and to the absorptive part of the elastic amplitude in the forward direction arises from the multi-Regge kinematics in the limit $s \to\infty$ in leading logarithmic approximation. In the next-to-leading logarithmic approximation, corrections require also the contribution from the quasi-multi-Regge kinematics ~\cite{FL98}. Hence, the subenergies between neighboring $s$-channel gluons must be large enough to remain in the Regge domain. At finite total energies, this implies that the amplitude is represented by a finite sum of $N$ terms~\cite{PR}, where $N$ increases like $\ln s$, rather than by the solution of the BFKL integral equation~\cite{FKL,BL,L}. The interest in the first few terms of the series is related to the fact that the energies reached by the present accelerators are not high enough to accommodate for a large number of $s$-channel gluons that eventually hadronize and give rise to clusters of secondary particles ~\cite{Lia}.
Consequently, one can expand the "supercritical" Pomeron $\sim s^{\alpha_p(0)}$ in powers of $\ln(s)$.

In Ref. ~\cite{PR} a model for the Pomeron at $t=0$
based on the idea of a finite sum of ladder diagrams in QCD was
suggested. According to the idea of that paper, the number of
$s$-channel gluon rungs and correspondingly the powers of
logarithms in the forward scattering amplitude depends on the
phase space (energy) available, i.e. as energy increases,
progressively new prongs with additional gluon rungs in the
$s$-channel open. Explicit expressions for the total cross section
involving two and three rungs or, alternatively, three and four
prongs (with $\ln^2(s)$ and $\ln^3(s)$ as highest terms,
respectively) were fitted to the proton-proton and
proton-antiproton total cross section data in the accelerator
region.
In a related paper ~\cite{Ducaty} the Pomeron was considered as a
finite series of ladder diagrams, including one gluon rung besides
the Low-Nussinov "Born term" ~\cite{LN} and resulting in a constant plus
logarithmic term in the total cross section. With a sub-leading
Regge term added, good fits to $pp$ and $p\bar p$ total as well as
differential cross section were obtained in ~\cite{Ducaty}.
There is  considered the opening channels (in $s$) as threshold
effects, the relevant prongs being separated in rapidity by $\ln
s_0$, $s_0$ being a parameter related to the average subenergy in
the ladder.
Within the "finite gluon ladder approach" to the Pomeron (see
~\cite{PR} and references therein), several options are possible.
In Ref. ~\cite{PR} a system of interconnected equations was
solved with several free parameters, including the value of
$s^i_0$, that determine the opening of each threshold (prong). In
that paper finite gluon ladders were calculated from QCD, where
the important dynamical information is
contained in $\rho$ of Eq. (7) of that paper, including $\ln s$
terms multiplied by the QCD running constant $\alpha_s$
constraining the interconnection between various powers of the
logarithms in the total cross section.
In \cite{PR2} an unitarization procedure was also included: the QCD-inspired amplitude was treated as a  Born term, subject to a subsequent unitarization procedure.

In present paper we use the abovementioned "finite gluon ladder approach" to the unitarized QCD-inspired amplitude of \cite{PR2} and calculate the values of $pp$ total cross section at LHC energy 7  and 14 TeV for several possible $s_0$.
For the sake of completeness, we {\bf compare} the $pp$ total cross section resulting from this model with those from popular alternatives, such as:  exponential, double logarithmic form and the DL two-Pomeron model \cite{DL}.
Fits to the $p\bar p$ and $pp$ data is performed from $\sqrt{s}=5 GeV$ up to highest energy data \cite{7tev}, \cite{sigtot}.

%Fits to the $p\bar p$ and $pp$ data were performed up to the highest energy Tevatron data $1.8$ $TeV$.

\section{Total cross sections from a finite sum of gluon ladders}

Following \cite{PR}, we write the Pomeron contribution to the total cross section
in the form
\begin{equation}
\sigma_t^{(P)}(s)=\sum_{i=0}^N
f_i\:\theta(s-s_0^i)\:\theta(s_0^{i+1}-s)\;, \label{z1}
\end{equation}
where
\begin{equation}
f_i=\sum^i_{j=0}a_{ij}L^j\;, \label{z2}
\end{equation}
$s_0$ is the prong threshold, $\theta(x)$ is the step function and
$L\equiv \ln(s)$. Here and in the following, by $s$ and $s_0$
respectively, $s$/(1\,GeV$^2$) and $s_0$/(1\,GeV$^2$) are implied.
The main assumption in Eq.~(\ref{z1}) is that the widths of the
rapidity gaps $\ln(s_0)$ are the same along the ladder. The
functions $f_i$ are polynomials in $L$ of degree $i$,
corresponding to finite gluon ladder diagrams in QCD, where each
power of the logarithm collects all the relevant diagrams. When
$s$ increases and reaches a new threshold, a new prong opens
adding a new power in $L$. In the energy region between two
neighboring thresholds, the corresponding $f_i$, given in
Eq.~(\ref{z1}), is supposed to represent adequately the total
cross section.

In Eq.~(\ref{z1}) the sum over $N$ is a finite one, since $N$ is
proportional to $\ln(s)$, where $s$ is the present squared c.m.
energy. Hence, this model is quite different from the "canonical"
approach \cite{FKL}, where, in the limit $s \to \infty$, the infinite sum of
the leading logarithmic contributions gives rise to an integral
equation for the amplitude.

To make the idea clear, we first describe the mechanism in the case of
three gaps (two rungs). In this case
\begin{equation}
f_0(s) = a_{00}\label{z3}
\end{equation}
for
\begin{equation}
\qquad s \leq s_0,\label{z4}
\end{equation}
\begin{equation}
f_1(s) = a_{10}+a_{11}L  \label{z5}
\end{equation}
for
\begin{equation}
\qquad s_0\leq s \leq s_0^2,\label{z6}
\end{equation}

\begin{equation}
f_2(s) = a_{20}+a_{21}L+a_{22}L^2  \label{z7}
\end{equation}
for
\begin{equation}
\qquad s_0^2\leq s \leq s_0^3,\label{z8}.
\end{equation}

By imposing the requirement of continuity (of the cross section
and of its first derivative) one constrains the parameters.

The same procedure can be repeated for any number of gaps.

Notice that the values of the parameters depend on the energy
range of the fitting procedure. For example, the values of the
parameters in $f_0$ is fitted in "its" range, i.e. for $s\leq
s_0$, will get modified in $f_1$ with the higher energy data and
correspondingly higher order diagrams included.

As a first attempt, only three rapidity gaps, that correspond to
two gluon rungs in the ladder were considered \cite{PR}. Fits to the $p\bar
p$ and $pp$ data were performed up to the highest energy Tevatron data $1.8$ $TeV$.  The value of the rapidity gap turned out to be  $\sqrt{s_0}\approx 12$ $GeV$, i.e. the value for which the energy range considered is covered with equal
rapidity gaps uniformly.

\section{Fit with the Unitarized QCD-inspired Pomeron}
% Explicit iterations of BFKL}

From the theoretical point of view, the phenomenological model of
Section 2 corresponds to the explicit evaluation in QCD of gluonic
ladders with an increasing number of $s$-channel gluons. This
correspondence is far from literal since each term of the BFKL
series takes into account only the dominant logarithm in the limit
$s\to \infty$. In the following we give concrete expressions for
the forward high energy scattering amplitudes for
hadrons in the form of an expansion in powers of large logarithms
in the leading logarithmic approximation.

We start from known results obtained in paper \cite{BL} where an
explicit expression for the total cross section for hadron-hadron
scattering has been obtained. In the high energy limit, it is
convenient to introduce the Mellin transform of the amplitude
\begin{equation}
%\begin{displaymath}
{\cal A}(\omega,t)=\int_0^\infty d \tilde s
\tilde s^{-\omega-1}\frac{\mbox{Im}_s {\cal A}(s,t)}{s},\
\tilde s=\frac{s}{m^2}.
%\end{displaymath}
\end{equation}  \label{z9}
% $$ s=\frac{s}{m^2}$$

To obtain the total cross section of proton-proton scattering it was  used ~\cite{PR2}
the ansatz of Ref.\cite{levin} for the impact factor of a hadron
in terms of its form factor $\Phi (q^2)$:
\begin{equation}\label{z15}
\Phi_{0}(k)=a k^2 e^{-bk^2}\;,
\end{equation}
as a result

\begin{displaymath}
\sigma_t(s)=\frac{\pi a^2}{2c} \left\{ 1+2 (\ln
2)\rho+\left[\frac{\pi^2}{12}+ 2 (\ln 2)^2 \right]\rho^2+ \right.
\end{displaymath}

\begin{equation} \label{u1}
\left. \frac{1}{3} \left[\frac{\pi^2}{2}(\ln 2)+4 (\ln
2)^3-\frac{3}{4}\zeta(3)\right] \rho^3+\ldots \right\}\;,
\end{equation}
where $\rho$ defined as

\begin{equation}\label{zr}
\rho=\frac{3\alpha_s}{\pi}\ln {\tilde s}
\end{equation}
and $\zeta(3)$ is the Riemann's Zeta function $\zeta(3)\approx1.202$. \\
This approach was used \cite{PR2} to predict the total cross section and ratio of real part of elastic amplitude to it imaginary part at LHC energy.
The strong coupling $\alpha_s$ is assumed to be frozen at a
suitable scale set, for example, by the external particles.

Now  we proceed with a fit and predictions with another Pomeron contribution corresponding to four rungs:

\begin{equation}\label{z11}
%\sigma_t(s)=\frac{\pi a^2}{2b} \left\{ 1+c_{1}\rho+c_{2}\rho^2+ %\left[\frac{\pi^2}{2c_{3}\rho^3+c_{4}\rho^4 +...
\sigma_t=(\pi a^2/(2b))[1+c_1\rho+c_2\rho^2+ (c_3/3)\rho^3+(c_4/12)\rho^4+...],
\end{equation}
\\where

\begin{displaymath}
c_{1}= {2}(\ln 2),     \qquad c_{2}=\frac{\pi^2}{12}+ 2 (\ln 2)^2,
\end{displaymath}

$c_{3} = 3,8511$ and  $c_{4} = 10,145$.

%\end{displaymath}

%$c_{3} = 3,8511.$ and $c_4=10,145$.\\

%%%%%%%%%%%%%%%%%%%%%%%%%%%%%%%%%%

Apart the calculations of the total cross section involving fits of a variety of Pomeron models, it is interesting to estimate the value of  $s_0$,
  a basic parameter in the finite series of QCD diagrams. Each set is 'active' in its rapidity gap, i.e. the parameters $a_{ij}$ in (\ref{z2}) should be fitted in each energy interval separately and relevant solutions should match by imposing continuity of the total cross section and of its first derivative. Here we present the result of a fits to the existing experimental data (including the new LHC data ~\cite{7tev}) and predict the values of the total cross section at the next LHC energy with the unitarized contribution of the  explicit BFKL iteration.
As will be seen below, the values of $s_0$ are quite constrained by the fits.

First of all, in the framework of our approach we mast put $\sqrt{s_0} > 5\ GeV$. As a convenient minimal value of rapidity gap we chosen $\sqrt{s_0} = 6.8 \ GeV$, where in estimating $\sigma_{tot}(pp)$ we are restricted by the maximal power $ln^4(s)$. From the experimental data  $\sqrt{s_0} \leq 24\ GeV$, one expects a high-quality description of the total cross section with the Pomeron contribution of Eq. (3), including relevant secondary Regge-pole terms. To cover the whole fitted range, we choose the interval $10.9\ GeV\leq \sqrt{s_0} \leq 24 \ GeV$,  where four gaps are sufficient to  estimate of total cross section at the maximal LHC energy.

 \begin{equation}
f_3(s) = a_{30}+a_{31}L+a_{32}L^2 +  a_{33}L^3 \label{s0 1}
\end{equation}
for
\begin{equation}
\qquad s_0^3\leq s \leq s_0^4\label{s0 2}.
\end{equation}
For the remaining part of the studied region
$6. 8 \;GeV\leq \sqrt{s_0} \leq 10.9\; GeV$ it is necessary to take into account five gaps to estimate the total cross section at the LHC.

\begin{equation}
f_4(s) = a_{40}+a_{41}L+a_{42}L^2 +  a_{43}L^3 + a_{44}L^4 \label{s0 3}.
\end{equation}
for
\begin{equation}
\qquad s _0^4\leq s \leq s_0^5. \label{s0 4}
\end{equation}

Below we perform such calculations in the framework of the eikonal formalism and
compare the results with the experimental data.
We start from Eq.(\ref{z11}) for the $pp$ and $\bar pp$ total
cross section. Supplying that expressing with an exponential
$t$-dependence, we get the elastic scattering amplitude:
\begin{displaymath}
F_{Born}(s,t)=A(-i\tilde s)^{1+\alpha't}\times
\end{displaymath}
\begin{equation}\label{amplitude}
[a_0+a_1\gamma\ln(-i \tilde s)+a_2\gamma^2\ln^2(-i
\tilde s)+a_3\gamma^3ln^3(-i \tilde s)]e^{Bt},
\end{equation}
where $\alpha'$ and $B$ are fitting parameters, $\gamma$ determined as (\ref{zr})
and
%%$$\gamma=\frac{3\alpha_s}{\pi,}$$
\begin{displaymath}
a_0=1+ \frac{\pi^2}{4}\Bigl(\frac{\pi^2}{12} + 2\ln^22\Bigr) \gamma^2,
\end{displaymath}
 \begin{displaymath}a_1=\frac{\pi^2}{4} \Bigl[\frac{\pi^2}{2}\ln2+ 4\ln ^32- \frac{3}{4}\zeta(3)\Bigr]\gamma^2 +2 \ln2,
 \end{displaymath}
%\Bigl(\frac{3\alpha_s}{2}\Bigr)^2[\frac{\pi^2}{2}\ln 2+4\ln^22-3\xi(3)/4],$$
\begin{displaymath}a_2=\frac{\pi^2}{12}+2\ln^22,$$
$$ a_3=\frac{1}{3}\Bigl[\frac{\pi^2}{2}\ln 2+4\ln^32-\frac{3}{4}\zeta(3)\Bigr].
\end{displaymath}
\begin{equation}\label {u3}
A=-\frac{a^2}{8c},\end{equation}

In the eikonalization procedure we follow Ref. \cite{Kroll},
according to which the Pomeron amplitude
\begin{equation}\label{Kroll}
F_P(s,t)=is\int_0^\infty
bdbJ_0\Bigl(b\sqrt{-t}\Bigr)\Bigl(1-e^{i\chi(b,s)}\Bigr),
\end{equation}
where $J_0$ is the Bessel function of zeroth order
%$p=\frac{1}{2m}\sqrt{s(s-4m^2)},$
 and the eikonal $\chi$ is
\begin{equation}\label{eikonal}
\chi(s,b)=\frac{1}{s}\int_0^{\infty}\sqrt{-t}d\sqrt{-t}I_0(b\sqrt{-t})F_{Born}(s,t).
\end{equation}

Inserting the expression for the Pomeron into Eq. (\ref{eikonal}) and expanding the exponential in (\ref{Kroll}), one find  for the eikonalized Pomeron amplitude

\begin{equation}\label{u4}
\label{series}F_P=2is\xi \sum_{k=1}^\infty\frac{1}{kk!}
\left(-\frac{\xi}{\mu}\right)^{k-1}e^{\mu t/k}.
\end{equation}

%%%%%%%%%%%%%%%%%%%%%%%%%%%%%%%%%%%%
\begin{figure}[h]
\begin{center}
\hspace{-1.cm}
\includegraphics[width=0.8\textwidth,angle=0]{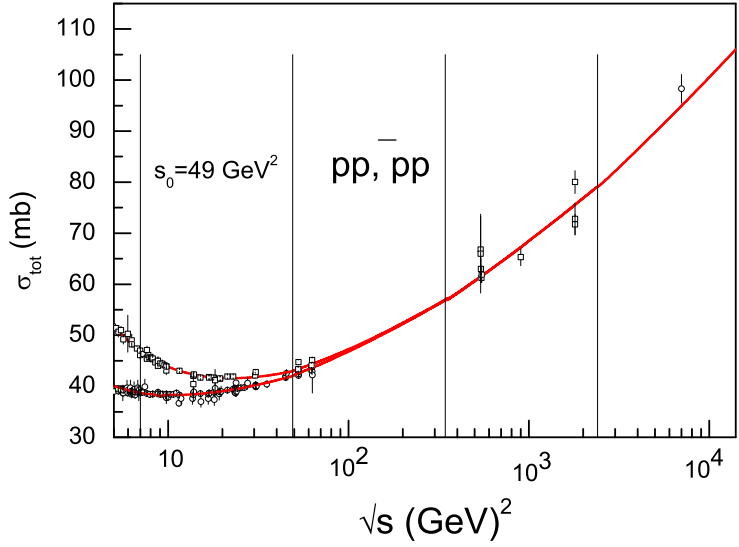}
\caption{\small \it {Total $\bar{p}p$ (upper curve) and $pp$ cross sections from the uniratized
(eikonalized) version of the model.The vertical lines corresponds to boundaries of rapidity gaps.}}
\end{center}
\end{figure}

The corresponding forward Pomeron
amplitude is
\begin{equation}\label{u5}
F_P(s,t=0)=2is\mu[C+ln(\xi/\mu)+E_1(\xi/\mu)],
\end{equation}\label{result}
where
\begin{equation}\label{u6}
\mu=B+\alpha'ln(-i \tilde s);\
\end{equation}
\begin{equation}\label{xi3}
\xi=\frac{A}{2m^2}(\xi_0+\xi_1+\xi_2+\xi_3),\ \
\end{equation}
 and
\begin{equation}\xi_0=a_0,\ \
\xi_1=a_1\gamma\ln(-i\tilde s),$$
$$\xi_2=a_2\gamma^2\ln^2(-i\tilde s),\ \
\xi_3=a_3\gamma^3\ln^3(-i\tilde s),
\end{equation}
where
%$$\gamma=\frac{2\alpha_s}{\pi},$$
 $C=0.577216 $
is the Euler constant and $E_1$ is the asymptotic form of the first order exponential integral:

\begin{equation}
E_1=\frac{\exp({-\xi/\mu})}{\xi/\mu} [1-\frac{1}{\xi/\mu}+
\frac{2}{(\xi/\mu)^2}-
\frac{6}{(\xi/\mu)^3}
 +....].
 \end{equation}

For the case of four rungs, the Born amplitude is

\begin{displaymath}
F_{Born}(s,t)=A(-i
\tilde s)^{1+\alpha't}[a_0+a_1\gamma\ln(-i \tilde s)+a_2\gamma^2\ln^2(-i
\tilde s)+
\end{displaymath}

\begin{equation}\label{u7}
+a_3\gamma^3ln^3(-i \tilde s)+
a_4\gamma^4ln^4(-i \tilde s)]e^{Bt},
\end{equation}
where

\begin{displaymath}
a_0=1+\left(\frac{\pi\gamma}{2}\right)^2\left(\frac{\pi^2}{12}+2\ln^2 2\right) + 5 c_4\left(\frac{\pi\gamma}{2}\right)^4,
\end{displaymath}

\begin{displaymath}
a_1=\frac{\pi^2}{4}\Bigl[\frac{\pi^2}{2}\ln 2+4\ln^32-\frac{3}{4}\zeta(3)\Bigr]\gamma^2+2ln2.
\end{displaymath}
\begin{displaymath}
a_2=\frac{3\pi^2}{2}c_4\gamma^2+\frac{\pi^2}{12}+2ln^{2}2,
\end{displaymath}
\begin{displaymath}
a_3=\frac{1}{3}\Bigl[\frac{\pi^2}{2}\ln 2+4\ln^32-\frac{3}{4}\zeta(3)\Bigr],
\end{displaymath}
\begin{displaymath}
a_4=\frac{1}{12}c_4\gamma^4
\end{displaymath}
%$$A=-\frac{a^2}{8c},$$
%$$\gamma=\frac{2\alpha_s}{\pi}$$
where $A$, $\mu$ and $\gamma$ are the same.
\begin{equation}\label {xi4}
\xi=\frac{A}{2m^2}(\xi_0+\xi_1+\xi_2+\xi_3 +\xi_4),
\end{equation}
and
\begin{displaymath}
\xi_0=a_0; \qquad
\xi_1=a_1\gamma\ln(-i\tilde s);\qquad
\xi_2=a_2\gamma^2\ln^2(-i\tilde s);
\end{displaymath}

\begin{displaymath}
\xi_3=a_3\gamma^3\ln^3(-i\tilde s);\qquad
\xi_4=a_4\gamma^4\ln^4(-i\tilde s).
\end{displaymath}

The obtained eikonalized Pomeron terms are appended by a
contributions from secondary reggeons, $\rho$ and $\omega$:

\begin{equation}\label{u11}
F_R^{\pm}(s,t=0)=g_f\tilde s^{\alpha_f(0)}\pm
ig_{\omega}\tilde s^{\alpha_\omega(0)},
\end{equation}
where the $+(-)$ sign
corresponds to $\bar pp(pp)$ scattering, the resulting forward
amplitude being
%$$F^{pp}_{\bar pp}=F^P(s,t=0)\pm F_R^{\pm}(s,t=0).$$
\begin{equation}\label{u12}
F^{\bar pp}_{pp}(s,t=0)=F_P(s,t=0)+ F_R^{\pm}(s,t=0).
\end{equation}
For the total cross section the norm
\begin{equation}\label{u13}
\sigma=\frac{4\pi}{s}Im F^{\bar pp}_{pp}(s,t=0)
\end{equation}
was used.
% and $\rho(s,0)=Re F^{\bar pp}_{pp}(s,t=0)/Im F^{\bar pp}_{pp}(s,t=0).$

Consider in more detail the fitting procedure in the approach with different numbers of rungs (power of log's).
In the case of 3 rungs, for values $\sqrt{s_0}$  within the interval
$10.9\ GeV$ -  $24\ GeV$ the whole range of the data
 $5 \ GeV \leq \sqrt{s}$ is covered.
 We  have chosen the parameters $B=0.116$ and $\alpha'=0.134$
 ~\cite{PR2}.

The parameters of secondary reggeons are the same for whole fitted experimental data region. Notice that in this approach for the Pomeron contribution there is only one free parameter in every separate interval. Therefore, for its determination, in principle, it is sufficient to have one experimental point. In our case this condition is fulfilled.

To estimate $s_0$ in the remaining part of the whole investigated region
$6,8 \ GeV \leq \sqrt{s_0}\leq 10.9 \ GeV $,  we added the new LHC \cite{7tev} data as well as the less reliable cosmic-ray data \cite{Honda}.
In our calculations $s_0$ is not a free parameter. Instead, we have performed  a series of fits and have chosen a value $s_0$ for the cases where
 $ \chi^2/dof \approx 1. $
We find, that this condition is valid in the whole interval
 $ 6.8 \ GeV \leq \sqrt{s_0} \leq 10.9 \ GeV $ ,  where $ \alpha_s $ is rather small $ \approx 0.08 $ and the prediction of total cross section at $7 \ TeV$ LHC energy crosses the experimental value within the error bars. In case three rungs, however, the predicted value is unacceptable low.
 In table 1. and fig 1. two of the best fits for $s_0 = 49 \ GeV^2$ and $s_0 = 121 \ GeV^2$ are quoted.
Throughout this paper we chosen the fits with ($\chi^2/dof)\leq 1.0$.

\begin{table}
\noindent\caption{Values of the fitted parameters and calculated total cross sections at the LHC energies.
$A$ in the (18)
%~(\ref{u2})
for different $i$ rung}\vskip3mm
\begin{center}{\small}
\begin{tabular}{|l|r|l||l|r|}
\hline
   &3 rungs && 4 rungs& \\
\hline
\hline
Parameter  & value & error & value & error \\
\hline

%&Parameter  & 3 rung & 4 rung &\\
  $A_0$ &  3.61 &  1.11 & 4.83 & 0.25   \\
\hline

 $A_1$ &  1.77 &  0.92 & 3.44 & 0.25   \\

 \hline

 $A_2$ &  1.06 &  0.64 & 2.63 & 0.26   \\
 \hline

 $A_3$ &  0.59 &  0.46 & 2.09 & 0.27   \\
 \hline

 $A_4$ &  - &  -       & 1.66 & 0.27   \\
\hline

  $\alpha_{s}$ & 0.145 & 0.049& 0.084 & 0.007\\
\hline
%  $B$ &  0.116 & & 0.116 &\\
%\hline
%$\alpha'$ & 0.134 &  & 0.134 &   \\
%\hline
 $g_{f}$ &  -6.17 & 0.53 & -7.67 & 0.32 \\
\hline
  $\alpha_{f}(0)$ &  0.681 & 0.028 & 0.633&0.011 \\
\hline

%$\alpha_{\omega}(0)$ &  0.46 & 0.46\\
$g_{\omega}$ &  3.61 & 0.17 & 3.72& 0.18\\
\hline
$\alpha_{\omega}(0)$ &  0.463 & 0.015 & 0.452 & 0.015 \\
%\hline

%$d$ &  1.77 & 1.45 & -0.48 & 0.98 \\
\hline
$s_0, GeV^2$ &  121.0 & - & 49.0 & - \\
\hline
$\sigma(pp),mb$&&&&\\
\hline
    $7  \ TeV $  &  92.2 & 1.5& 94.8 & 1.5\\
    $14 \ TeV$  &  102.6& 2.1& 106.6& 1.9\\
\hline
\hline
\end{tabular}
\end{center}
\end{table}

%Everywhere in this paper we chosen the fits with ($\chi^2/dof)\leq 1.0$.

\section{$\sigma_{tot}(s)$ fits for different Pomeron models}

Here we repeat our previous fit \cite{PR2} limiting ourselves by $163$ data points on $pp$ and $\bar pp$ total cross-sections in the range $5  \ GeV \leq\sqrt{s}\leq 1.8 \ TeV$ ~\cite{sigtot}, however adding the new LHC experimental point at $7 \ TeV$ with the results and predictions for $pp$ total cross section at the  LHC energies
quoted in Table 2.
For the completeness and comparison we performed also the fits of known Regge-models with different the Pomerons used.
The variety of the  Pomeron-models \cite{Ezhela} for the total cross section, can be distributed to three groups:

%Dipole Pomeron
%5\begin {equation}\label{r1}
%Im F_p\left(s,t=0\right)=a_0+a_1 ln\left(\tilde{s}\right),\\
%\tilde{s}=s/s_0,
%\end {equation}
Tripole Pomeron
\begin {equation}\label{r2}
Im F_p\left(s,t=0\right)=a_0+a_1ln^2\left(\tilde{s}\right),
\end {equation}
Supercritical soft Pomeron
\begin {equation}\label{r3}
Im F_p\left(s,t=0\right)=a_0+a_1 \tilde{s}^{\alpha_s}.
\end {equation}
To this end we have added the two-component Donnachie-Landshoff Pomeron model ~\cite{DL}
\begin {equation}\label{r4}
Im F_p\left(s,t=0\right)=a_0+a_1 \tilde{s}^{\alpha_s}+ a_2 \tilde{s}^{\alpha_h}.
\end {equation}

The hard Pomeron contribution to the $pp$- and $\bar{p}p$- total cross section is much smaller than the soft one and it is hard to determine it
directly.
However, as it was shown in \cite{Cudell}, the data for the total cross section and for real part of elastic amplitude indicate  the presence of a hard Pomeron in $\pi$$ p$ and $Kp$ elastic scattering at $t=0$, compatible with that observed in deep inelastic scattering. The inclusion of these two Pomerons, together with use of the integral dispersion relation and subleading Reggeons, leads to a successful description of all hadron and photon scattering on proton data for $5 \ GeV \leq s \leq 100 \ GeV$ \cite{Cudell}. The two-Pomeron  approach is also compatible with $pp$ and $ \bar p p$ data, provided one unitarizes it at high energies. It was  found that the contribution of the hard Pomeron relative to the soft one is $0.17/55$ (see Tab. 4. in ~\cite{Cudell}).
In the last case we eikonalized the Pomeron terms (see the previous part).

For the soft and hard Pomeron, the eikonalized contribution  has the form \label {xi}

$$\xi=\xi_0+\xi_1+\xi_2,$$
with
$$\xi_0=a_0,\ \
\xi_1=a_1\left(-\tilde s\right)^{\delta_s},$$
corresponding to to the soft Pomeron and
$$\xi_2=a_2\left(-\tilde s\right)^{\delta_h}$$
 corresponding to the hard one, where $\delta_s=0.08$ and $\delta_h = 0.45$ ~\cite{Cudell}.
To perform the fit we need to include the contributions from secondary reggeons, $\rho$ and $\omega$ ~(\ref{u11}).
The normalization of the total cross section is the same as in the previous secttion ~(\ref{u13}).
Fits of all models to $pp$ and $\bar{p}p$ total cross section data  were performed in the range $5 \ GeV \leq\sqrt{s}\leq 7$ TeV, the quality of fit being $\chi^2/dof\leq 1.0$. The results along with the expected values for the  total cross section at maximal LHC energy are quoted in Table 2.

\begin{table}
\noindent\caption{Prediction of Pomeron models for total cross section at LHC energies}\vskip3mm
%\begin{center}{\small
\begin{center}{\small\begin{tabular}{|l|c|c|}
\hline
 & $\sigma_{tot}\left(pp\right),mb$ &  \\
\hline
$\sqrt{s}, TeV $& 7 & 14   \\

\hline

 tripole    &  98.4  $\pm$ 0.9& 110.4$\pm$ 0.7  \\
\hline
  soft      &  98.7 $ \pm$0.9 & 111.8$\pm $1.0\\
\hline
  sof+hard  &  101.8 $ \pm $ 1.4 & 120.1 $\pm$ 1.7\\
\hline
QCD, 2 rungs&  96.5  $\pm$0.5 & 107.6 $\pm$ 0.6\\
\hline
QCD, 3 rungs &  94.9 $ \pm$1.3 & 105.7 $\pm$ 1.3\\
\hline
 QCD, 4 rungs &  96.8 $ \pm $ 1.3 & 107.1 $\pm$ 1.5\\
\hline
\end{tabular}}
\end{center}
\end{table}

All calculated values fit the new experimental data ~\cite{7tev} and predict almost the same values at next LHC energy except  the unitarized two-pomeron model, which  predict the value exceeding $120$ mb.

\begin{center}

\end{center}
%\bf E'RTHETETLEN: on that none of the conventional models no more takes place and in soft processes  %presents the hard Pomeron}.

\section{Conclusions}

Our main goal was an adequate picture of the Pomeron
exchange at $t=0$.  In our opinion, it is neither an infinite sum
of gluon ladders as in the BFKL approach~\cite{FKL,BL,L}, nor its
power expansion. In fact, the finite series - call it "threshold
approach" - considered in Sec. 2. and 4. and in the previous
papers~\cite{PR}, realizes a non-trivial dynamical balance between
the total reaction energy and the subenergies equally partitioned
between the multiperipheral ladders.

The role and the value of the width of the gap $s_0$, is an
important physical parameter, independent of the
model presented above. We have fitted it and obtained the value which exceeds our previous estimates ~\cite{PR}. \
Another goal of the present investigation was the prediction of the proton-proton total cross section at the maximal LHC energy with several  Regge- and QCD-inspired Pomeron models. We checked the traditional Regge-models as well as the controversial case of the hard Pomeron. In this case we predict the largest value for the total cross section, $ \approx 120\ mb $ (see also ~\cite{selugin}). All model fits give values of $\sigma\left(s\right)$ compatible with the experiment at $7$ TeV within the error bars and consequently cannot be ruled out.

\section*{Acknowledgments}

This work was done in the framework of the Hungarian Academy of Sciences Fellowship for Transfrontiery Hungarian Science.
We thank V.~Bitev, L.~Jenkovszky, and E.~Kuraev and for discussions and valuable collaboration.

\end{document}